\RequirePackage{tikz}

\documentclass[12pt]{article}

\usepackage{graphicx}%
\usepackage{multirow}%
\usepackage{amsmath,amssymb,amsfonts}%
\usepackage{amsthm}%
\usepackage{mathrsfs}%
\usepackage[title]{appendix}%
\usepackage{xcolor}%
\usepackage{relsize} 
\usepackage{cite}

\theoremstyle{thmstyleone}%

\theoremstyle{thmstyletwo}%
\newtheorem{example}{Example}%
\newtheorem{remark}{Remark}%

\theoremstyle{thmstylethree}%

\usepackage{upgreek,bm}%

\newcommand{\Frac}[2]{\mathchoice{\text{\small$\frac{#1}{#2}$}}
                                 {\text{\large$\frac{#1}{#2}$}}
                                 {\text{\large$\frac{#1}{#2}$}}
                                 {\text{\large$\frac{#1}{#2}$}}}
\newcommand{\eq}[1]{(\ref{#1})}

\newcommand{\pl}{\partial}

\def\vv{\bm v}

\def\uu{\bm u}

\def\ee{\bm e}
\def\FF{{\bm F}}

\def\bbD{\mathbb D}

\def\bbI{\mathbb I}

\newcommand{\lineunder}[2]{\LU{\begin{array}[t]{c}\underbrace{#1}\vspace*{.5em}\end{array}}{\mbox{\footnotesize\rm #2}}}
\newcommand{\LU}[2]{\begin{array}[t]{c}#1\vspace*{-1em}\\_{#2}\end{array}}

\usepackage{psfrag}    
\usepackage{tikz}  
\usetikzlibrary{decorations.pathmorphing}
\usetikzlibrary{patterns}
\usepackage{xcolor} 
\usepackage{pgfplots}
\usetikzlibrary{calc}
\newcommand{\R}{{\mathbb R}}

\def\DTee{\bm\varepsilon}
\newcommand{\SUM}{\text{{\small$\displaystyle{\sum}$}}}
\newcommand{\Sum}{\text{{\footnotesize$\displaystyle{\sum}$}}}
\newcommand{\Mid}{{\mid}}
\newcommand{\zetap}{\zeta_{\rm vp}}
\def\strain{\bm e}

\begin{document}

\noindent
{\LARGE\bf A few notes about viscoplastic rheologies}

\bigskip\bigskip\bigskip

\noindent
{\large Tom\'a\v s Roub\'\i\v cek}

\bigskip\bigskip\bigskip

{\small
\noindent
    {\bf Abstract}.
The rigorous tools of convex analysis are used to examine
    various serial and parallel combinations of
    linear viscosity and perfect plasticity. Nonlinear
    viscosities are also considered. The general aim
    is to synthesize a single convex ``viscoplastic''
    dissipation potential from the potentials of particular
    viscous or plastic elements. Rigorous serial-viscosity
    models are then compared with empirical models
    based on harmonic means, which are commonly used
    for various geomaterials.

\medskip

\noindent {\bf AMS Subject Classification.} 
  74C10, 
  74L05, 
  76A10, 
86A04. 

 \medskip

\noindent {\bf Keywords}.
dissipation potentials, convex conjugate, infimal convolution,
  Bingham fluids, shear-thinning non-Newtonian fluids, Norton-Hoff model,
  power-law fluids, rheology of rocks and ice.

}

\bigskip\bigskip

\section{Introduction}\label{sec1}
There is a menagerie of viscous or viscoplastic models for non-Newtonian fluids. These
are also used for various viscoelastic materials in the shear parts, combined with the
solid-type rheologies in the volumetric parts, which  yields advanced models
of creep, relaxation, and plasticity in many engineering and geophysical materials.

Without substantially restricting the applicability, we confine ourselves to situations
in which the viscoplastic rheology is governed by a convex (pseudo)potential, denoted by
$\zetap$, in the sense that the dissipative stress $\bm\sigma$ equals to the derivative
of $\zetap$ depending here on the {\it strain rate} $\DTee$, not on the strain itself.
Specifically,
\begin{align}\label{...}
  \bm\sigma=\zetap'(\DTee)\ \ \ \text{ or, in a certain detail, }\ \
  \bm\sigma=
  \bm\mu_{\rm eff}^{}\DTee\ \ \text{ with }\ \bm\mu_{\rm eff}^{}=\bm\mu_{\rm eff}^{}(\DTee)\,,
\end{align}
where $\bm\mu_{\rm eff}^{}$ is in the position of the so-called {\it effective}
(also called {\it apparent}) {\it viscosity} (in SI unit
Pa${\cdot}$s\,=\,J\,s/m$^3$=\,kg${\cdot}$m$^2$/s) depending generally on $\DTee$,
here obviously as $\bm\mu_{\rm eff}^{}(\DTee)=
\zetap'(\DTee)\DTee^{-1}$; noteworthy, this formula works for regular
(i.e.\ invertible) matrices $\DTee$, otherwise $\bm\mu_{\rm eff}^{}(\cdot)$
can be defined by continuity if the limit for an irregular $\DTee$ exists.
This can be applied to the small-strains or also to various large-strain models where
various objective strain rates can be used in the place of $\DTee$. Similarly, various
combinations with elasticity that lead to visco-elastoplastic rheologies can be considered.
Focusing solely on the viscous or viscoplastic rheologies, we intentionally avoid being
specific about such extensions beside Remark~\ref{rem-conjug-visco} below.

The plan of this paper is as follows. First, in Section~\ref{sec-viscoplasticity},
we discuss two basic scenarios for merging the  (perfectly) plastic and the linear
viscous phenomena: parallel and serial. Then, in Section~\ref{sec-viscoplasticity-2},
we examine two combinations of one plastic and two viscous elements as used in
literature, showing their mutual equivalence. Finally, in Section~\ref{sec-nonlin-visco},
we discuss modifications involving nonlinear power-law viscosities, which are often used
in applications. For the readers' convenience, the basic definitions from convex analysis
are summarized in the Appendix.

 We will use the general notational convention that tensors or vectors
(or tensor- or vector-valued fields) are denoted as boldfaced, while 
scalars or scalar-valued fields are in normal Italics or Greek fonts.
Moreover, $(\cdot)'$ will denote the derivative of a differentiable functional
while $\pl(\cdot)$ will denote the generalized derivative of
a non-differentiable convex functional, cf.\ \eq{app:subdifferential} below.
Actually, \eq{...} could have also been considered more generally for
$\zetap$ non-differentiable at $\DTee=\bm0$ written as an inclusion
$\bm\sigma\in\pl\zetap(\DTee)$.

\section{Viscoplasticity: basic scenarios}\label{sec-viscoplasticity}
\begin{figure}
\begin{center}
\psfrag{z}{\small $\DTee$}
\psfrag{z1}{\small $\DTee_1$}
\psfrag{z2}{\small $\DTee_2$}
\psfrag{s1}{\small $\bm\sigma_1$}
\psfrag{s2}{\small $\bm\sigma_2$}
\psfrag{d}{\small $D$}
\psfrag{sA}{\small $\sigma_\text{\!\sc a}^{}$}
\psfrag{perfect plastic}
{\footnotesize \begin{minipage}[t]{16em}\hspace*{0em}a perfectly plastic (of a dry-friction
\\[-.1em]\hspace*{9.5em}type) element\end{minipage}}
\hspace{1em}\includegraphics[width=28em]{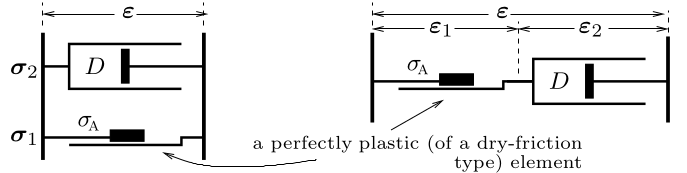}
\end{center}
\vspace*{-.2em}
\caption{
{\sl Two options in combination of a viscous damper with
  a perfect{-}plasticity  element  with the activation
  threshold $\sigma_\text{\!\sc a}^{}$.
}}
\label{fig-visco-plastic}
\end{figure}

First, we consider the perfect plasticity governed by the convex  positively 
homogeneous  of  degree 1 (non-smooth) potential
$\zeta_1(\cdot)=\sigma_\text{\!\sc a}^{}\Mid\cdot\Mid$ with
$\sigma_\text{\!\sc a}^{}>0$  denoting  an {\it activation} (yield) {\it stress} in
Pa=J/m$^3$ and a linear  Newton-type (sometimes called Stokes-type) 
viscosity governed by the homogeneous degree-2 (here
quadratic) potential $\zeta_2(\cdot)=\frac12D\Mid\cdot\Mid^2$ with $D>0$ a
viscosity coefficient in Pa${\cdot}$s. There are two basic options to combine
the plastic-like element governed by the degree-1 potential with the creep-like,
rate-dependent, linearly viscous element governed by the degree-2 (quadratic)
potential, as depicted in Figure~\ref{fig-visco-plastic}.

The first option, depicted sche\-ma\-ti\-cally on
Figure~\ref{fig-visco-plastic}-left, leads to
\begin{align}\nonumber
  \bm\sigma=\bm\sigma_1+\bm\sigma_2\ \ \text{ with }\ \
  \bm\sigma_1\in\pl\zeta_1(\DTee)\ \text{ and }\ \bm\sigma_2=\zeta_2'(\DTee),\ \text{  i.e. }\
\bm\sigma\in\pl\zetap(\DTee)\
\\\text{ with }\ \ \zetap=\zeta_1+\zeta_2
  \ \ \text{ for }\ \ \zeta_1(\cdot)=\sigma_\text{\!\sc a}^{}\Mid\cdot\Mid\ \text{ and }\
\zeta_2(\cdot)=\mbox{$\frac12$}D\Mid\cdot\Mid^2;
\label{GSM-Jeffreys-plast}\end{align}
note that $\zeta_1$ and thus also $\zetap$ are non-differentiable at $\DTee=0$
so that we have used the subdifferential ``$\,\pl\,$'' as a generalization of
the derivative, cf.\ \eq{app:subdifferential} below, and the inclusion
``$\,\in\,$'' instead of an equality. This model composes the linear creep and
the perfect plasticity in parallel. In engineering, such model is also known as
{\it rate-dependent plasticity} or {\it Kelvin viscoplasticity}
\cite{DuBoPo19FTSB} or, most often, as the {\it Bingham fluid}.
\begin{figure}
\begin{center}
\psfrag{potential0}{\small $\zetap$}
\psfrag{potential}{\small $\zetap=\zeta_1^{}\!+\zeta_2^{}$}
\psfrag{potential1}{\small $\zeta_1^{}$}
\psfrag{potential2}{\small $\zeta_2^{}$}
\psfrag{yield}{\small $\sigma_\text{\!\sc a}^{}$}
\psfrag{Dz}{\small $\DTee$}
\psfrag{slope}{\footnotesize slope $D$}
\psfrag{1/slope2}{\footnotesize slope $1/D$}
\psfrag{subdifferential}{\small $\pl\zetap$}
\psfrag{conjugate}{\small$\hspace*{-6.3em}{\zetap^*}\!\!'=[\pl\zetap]^{-1}$}
\psfrag{conjugate-potential}{\small $\!\zetap^*$}
\psfrag{s}{\small $\bm\sigma$}
\psfrag{0}{\small $0$}
\hspace*{1.5em}\includegraphics[width=32em]{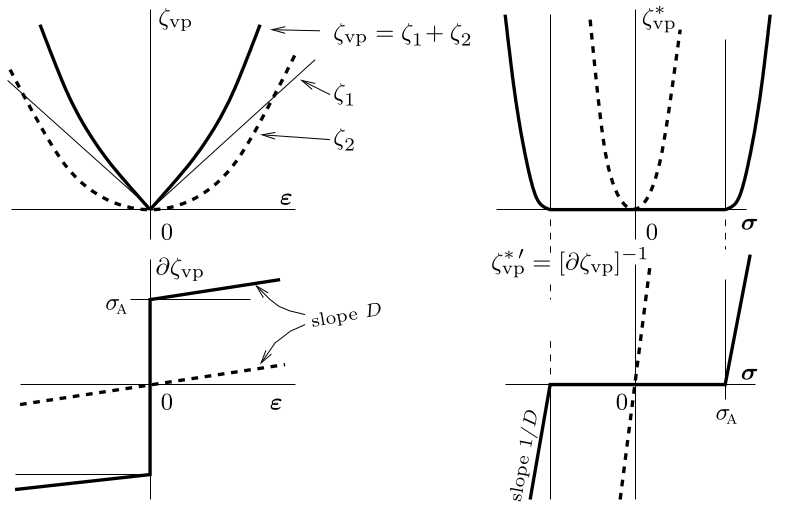}
\end{center}
\vspace*{-.6em}
\caption{
{\sl Schematic illustration of the convex nonsmooth dissipation potential
$\zetap$ from \eq{GSM-Jeffreys-plast} with both $D_2>0$ and
$\sigma_\text{\!\sc a}^{}\!>0$, its subdifferential $\pl\zetap$ which
is set-valued at 0, and its (smooth) conjugate $\zetap^*$ and its
derivative (=\,single-valued inverse of $\pl\zetap$) used to model
(rate-dependent) plasticity; cf.\ also \cite[Chap.27]{JirBaz02IAS}.}}
\label{fig-plastic}
\end{figure}

\begin{figure}
\begin{center}
\psfrag{potential0}{\small $\zetap$}
\psfrag{potential}{\small $\zetap=\zeta_1^{}\Box\,\zeta_2^{}$}
\psfrag{potential1}{\small $\zeta_1^{}$}
\psfrag{potential2}{\small $\zeta_2^{}$}
\psfrag{potential1*}{\small $\zeta_1^*$}
\psfrag{potential2*}{\small $\zeta_2^*$}
\psfrag{yield}{\small $\sigma_\text{\!\sc a}^{}$}
\psfrag{Dz}{\small $\DTee$}
\psfrag{subdifferential}{\small $\zetap'$}
\psfrag{slope}{\footnotesize  slope $D$}
\psfrag{1/slope2}{\footnotesize \!\!\!slope $1/D$}
\psfrag{conjugate}{\small$\hspace*{-6.2em}\pl\zetap^*=[\zetap']^{-1}$}
\psfrag{conjugate-potential}{\small $\zetap^*$}
\psfrag{s}{\small $\bm\sigma$}
\psfrag{0}{\small $0$}
\psfrag{aseismic region}
{\footnotesize\sf\begin{minipage}[t]{10em}\hspace*{0em}creep 
\\[-.4em]\hspace*{0em}region\end{minipage}}
\hspace*{.5em}\includegraphics[width=33em]{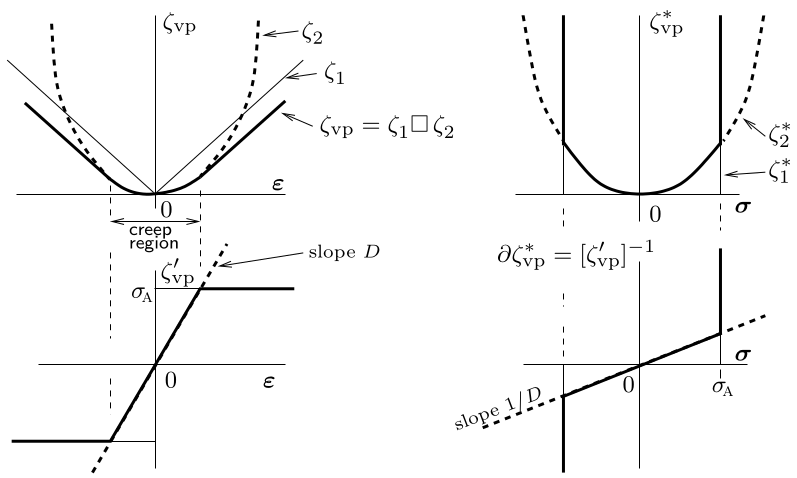}
\end{center}
\vspace*{-.4em}
\caption{
{\sl Schematic illustration of the convex smooth dissipation potential
$\zetap$ from \eq{GSM-Jeffreys-plast+} with both $D_2>0$ and
$\sigma_\text{\!\sc a}^{}\!>0$, its continuous differential $\zetap'$,
and its (nonsmooth) conjugate $\zetap^*$ and its 
derivative (=\,set-valued inverse of $\zetap'$) used to model
plasticity combined with a creep.}}
\label{fig-plastic-in-series}
\end{figure}

An ``opposite'' model is the composition of the linear creep and
the prefect plasticity in series, cf.\ Figure~\ref{fig-visco-plastic}-right,
which is sometimes referred to as viscoplasticity in a narrower sense.
To identify the dissipative response of this configuration,
let us decompose the total strain rate $\DTee$ into a 
sum of strains governed by a common stress, i.e.\ $\DTee=\DTee_1^{}\!+\DTee_2^{}$ with
$\zeta_1'(\DTee_1^{})=\bm\sigma=\zeta_2'(\DTee_2^{})$.
Exploiting the convex-conjugation operator $(\cdot)^*$ defined by \eq{app:Legendre-Fenchel}
below, the notation for the derivative $(\cdot)'$, and the formula \eq{e4.Fenchel-equiv}, we can
write $\DTee_1^{}={\zeta_1^*}'(\bm\sigma)$ and $\DTee_2^{}={\zeta_2^*}'(\bm\sigma)$,
where we used the brief notation ${(\cdot)^*}'=((\cdot)^*)'$. Furthermore,
using the infimal-convolution ``\,$\Box$\,'' defined by \eq{infimal-convolution} below,  
we can merge both dissipative potentials $\zeta_1^{}$ and $\zeta_2^{}$ into a
single dissipation potential $\zetap=\zeta_1^{}\Box\,\zeta_2^{}$, more specifically we use the
formula \eq{infimal-convolution-conjugate} for 
\begin{align}\label{GSM-Jeffreys-plast+}
\DTee={\zeta_1^*}'(\bm\sigma)+{\zeta_2^*}'(\bm\sigma)
=\big[\zeta_1^*{+}\,\zeta_2^*\big]'(\bm\sigma)
={\zetap^*}\!\!'(\bm\sigma)\ \ \text{ with }\ 
\zetap=\zeta_1^{}\Box\,\zeta_2^{}\,.
\end{align}
The same formula holds if one of these potentials is not smooth which, however,
yields beneficially a smooth potential $\zetap$, cf.\
Figure~\ref{fig-plastic-in-series}. This alternative model is particularly suitable
for merging  {\it creep} under low driving stresses and a {\it fast slip} when
the driving stresses exceed a prescribed threshold. The creep occurs even under
very low stresses smoothly in very long timescales and may lead to very
large displacements without triggering any fast ``catastrophic'' events like
earthquakes.
It should be remarked that an explicit form for $\zeta_1^{}\Box\,\zeta_2^{}$
can be not trivial in particular cases. Interestingly, for the quadratic
function $\zeta_2^{}$ as in \eq{GSM-Jeffreys-plast}, $\zeta_1^{}\Box\,\zeta_2^{}$
is the {\it Yosida approximation} of $\zeta_1^{}$, namely $\mathscr{Y}_{1/D}\zeta_1$,
cf.\ \eq{Yosida-approx} below. In this case, $\zetap$ is smooth
(continuously differentiable). For a 1-homogeneous $\zeta_1^{}$ as in
\eq{GSM-Jeffreys-plast}, the explicit form of $\zeta_1^{}\Box\,\zeta_2^{}$
is known as a so-called Huber function \cite{Hube81RS}, cf.\ also
\cite{BecTeb12SFOM}, i.e.\ here
\begin{align*}
\big[\zeta_1^{}\Box\,\zeta_2^{}\big](\DTee)=\begin{cases}
\frac12D\Mid\DTee\Mid^2&\text{for }\ \Mid\DTee\Mid\le\sigma_\text{\!\sc a}^{}/D\ \ \
\text{ (i.e.\ the creep regime)}\,,
\\\sigma_\text{\!\sc a}^{}\Mid\DTee\Mid-\frac12\sigma_\text{\!\sc a}^2/D\!\!
&\text{for }\ \Mid\DTee\Mid>\sigma_\text{\!\sc a}^{}/D
\ \ \ \text{ (i.e.\ a fast-slip regime)}\,.
\end{cases}
\end{align*}
The convex conjugate $[\zeta_1^{}\Box\,\zeta_2^{}]^*$ is then quadratic
for $\Mid\bm\sigma\Mid\le\sigma_\text{\!\sc a}^{}$ in the  creep (in geophysics
called ``aseismic'')  regime, specifically
$[\zeta_1^{}\Box\,\zeta_2^{}]^*(\bm\sigma)=\frac12D^{-1}\Mid\bm\sigma\Mid^2$,
otherwise $=+\infty$, as in Figure~\ref{fig-plastic-in-series}-right-up; this
function was used also in \cite{ChLaTh25ACHM,EiHoMi22LHSV}.
A certain analytical drawback of this 
is the discontinuity of ${\zetap^*}\!\!'$ or, more precisely of the
set-valued subdifferential $\pl\zetap^*$ of the convex conjugate $\zetap^*$,
cf.\ Figure~\ref{fig-plastic-in-series}-right-down.

In general, the viscoplastic response is often described by the so-called
{\it effective} (also called {\it apparent}) {\it viscosity}
$\mu_{\rm eff}^{}\!=\mu_{\rm eff}^{}(\DTee)$ (in SI unit Pa${\cdot}$s)
such that $\bm\sigma=\mu_{\rm eff}^{}\DTee$, cf.\ \eq{...}. Here this means, for
$\DTee\ne\bm0$, that
\begin{align}
\hspace*{-1em}\mu_{\rm eff}^{}(\DTee)=\begin{cases}
\min\big(D,\Frac{\sigma_\text{\!\sc a}^{}}{\Mid\DTee\Mid}\big)\!\!\!\!
&\text{for the visco-perfectly-plastic model, Fig.\,\ref{fig-visco-plastic}-right},
\\D+\Frac{\sigma_\text{\!\sc a}^{}}{\Mid\DTee\Mid}&\text{for
the Bingham-type model, Fig.\,\ref{fig-visco-plastic}-left}\,.
\end{cases}
\nonumber\\[-2em]
\label{effect-visco}\end{align}
For the min-formula in \eq{effect-visco} see e.g.\
\cite{ArGoHu17SSTZ,Gler18NVAB,Kara08DEMI}.

When combined with the elastic element in series,
the parallel viscoplastic model from Figure~\ref{fig-visco-plastic}-left
is sometimes referred as a {\it Bingham model} \cite[Chap.27]{JirBaz02IAS}
while the serial viscoplastic model from Figure~\ref{fig-visco-plastic}-right
referred as an {\it extended Maxwell model} \cite{Buro11RSL}.

\begin{remark}[{\sl Generalized Bingham-type viscoplasticity.}]\label{rem-Binghamo}\upshape
  The conjugate potential $\zetap^*$ from Figure~\ref{fig-plastic} can be
  expressed as $\zetap^*(\bm\sigma)=\phi(\max(0,f(\bm\sigma)))$ with a
  ``flow function'' $\phi(\cdot)=\frac12D^{-1}\Mid\cdot\Mid^2$ and an
  ``overstress'' $f(\cdot)=\cdot-\sigma_\text{\!\sc a}^{}$. Then
$\DTee={\zetap^*}\!\!'\,(\bm\sigma)=\phi'(\max(0,f(\bm\sigma)))f'(\bm\sigma)$. 
  In the engineering literature, this form is a departure for many
  generalizations. In particular, a general polynomial
  $\phi(\cdot)=D^{-1}\Mid\cdot\Mid^{1+n}/(1{+}n)$ with some $n>0$ is used, cf.\
  also \cite[Chap.27]{JirBaz02IAS}.
\end{remark}

\section{Viscoplasticity: three-element scenarios}\label{sec-viscoplasticity-2}
\begin{figure}
\begin{center}
\psfrag{z}{\small $\DTee$}
\psfrag{z2}{\small $\DTee_2$}
\psfrag{z1}{\small $\DTee_1$}
\psfrag{d2}{\small $D_2$}
\psfrag{d3}{\small $D_3$}
\psfrag{d2'}{\small $\widetilde D_2$}
\psfrag{d3'}{\small $\widetilde D_3$}
\psfrag{sA}{\small $\sigma_\text{\!\sc a}^{}$}
\psfrag{sA'}{\small $\widetilde\sigma_\text{\!\sc a}^{}$}
\psfrag{s}{\small $\bm\sigma$}
\hspace{.0em}\includegraphics[width=27em]{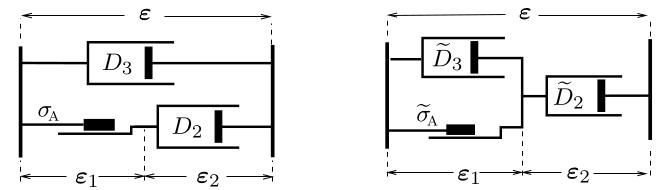}
\end{center}
\vspace*{-.2em}
\caption{
{\sl Two (mutually equivalent) variants of a combination of two
viscous dampers with a perfect plasticity with the activation thresholds
$\sigma_\text{\!\sc a}^{}$ and $\widetilde\sigma_\text{\!\sc a}^{}$ leading to a
continuous ${\zetap^*}\!\!'$.
}}
\label{fig-visco-plastic-1}
\end{figure}

In addition, the two options from Figure~\ref{fig-visco-plastic} can be further
combined with linear  Newton-type viscosity.
This involves two linear viscosities, sometimes called a {\it bi-viscosity model}
\cite[Sect.4.4]{Huil15FMV}, and thus devises more realistic and even mathematically simpler
models with more parameters and easier to be implemented numerically.
These two variants are illustrated in Figure~\ref{fig-visco-plastic-1}.

\begin{figure}
\begin{center}
\psfrag{sA}{\small $\sigma_\text{\!\sc a}^{}$}
\psfrag{potential0}{\small $\zetap$}
\psfrag{potential}{\footnotesize $\zetap\!=(\zeta_1^{}\Box\,\zeta_2^{})+\zeta_3^{}$}
\psfrag{potential1}{\small $\zeta_1^{}$}
\psfrag{potential2}{\small $\zeta_2^{}$}
\psfrag{potential3}{\small $\zeta_3^{}$}
\psfrag{gradient}{\small $\!\!\!\zetap'$}
\psfrag{slope3'}{\footnotesize slope  $1/D_3^{}$}
\psfrag{slope23'}{\footnotesize slope  $1/(D_2^{}{+}D_3^{})$}
\psfrag{slope3}{\footnotesize slope  $D_3^{}$}
\psfrag{slope23}{\footnotesize slope  $D_2^{}{+}D_3^{}$}
\psfrag{yield}{\footnotesize $\sigma_\text{\!\sc a}^{}(1+\frac{D_3}{D_2})$}
\psfrag{s}{\small $\bm\sigma$}
\psfrag{switch}{\footnotesize $\sigma_\text{\!\sc a}^{}/D_2^{}$}
\psfrag{conjugate}{\small$\hspace*{2.5em}{\zetap^*}\!\!'$}
\psfrag{Dz}{\small $\DTee$}
\hspace{-.5em}\includegraphics[width=38em, height=11em]{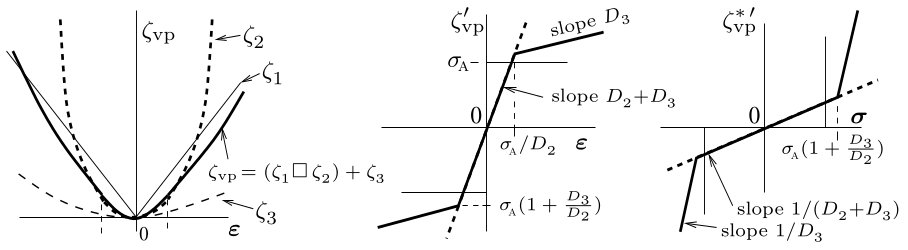}
\end{center}
\vspace*{-.2em}
\caption{
{\sl Schematic illustration of the bi-visco-plastic model from 
from Figure~\ref{fig-visco-plastic-1}-left
leading to a continuously differentiable ${\zetap^*}\!\!'$ with
${\zetap^*}$ continuously differentiable.
}}
\label{fig-visco-plastic-2}
\end{figure}

The first option, as depicted in Figure~\ref{fig-visco-plastic-1}-left, copes
with the mentioned drawback of the set-valued subdifferential of $\zetap^*$
in the serial viscoplastic model from Figure~\ref{fig-visco-plastic}-right. This
leads to the strictly convex piecewise-$C^1$ potential
\begin{align}\label{GSM-Jeffreys-plast++}
\zetap=\big(\zeta_1^{}\Box\,\zeta_2^{}\big)+\zeta_3^{}
\ \ \ \ \text{ with }\ \ \zeta_i^{}(\cdot)={\Frac12}D_i^{}\Mid\cdot\Mid^2,\ \ i=2,3,
\end{align}
and with $\zeta_1^{}$ as before. More in detail, realizing the modification of
\eq{GSM-Jeffreys-plast+} as
$\DTee=[\zeta_1^*{+}\,\zeta_2^*]'(\bm\sigma-D_3\DTee)$, written as
$[\zeta_1\Box\,\zeta_2]'(\DTee)+D_3\DTee=\bm\sigma$, we obtain
the equation $\zetap'(\DTee)=\bm\sigma$ with
$\zetap$ identified as in \eq{GSM-Jeffreys-plast++}.
Such differentiable $\zetap^{}$ together with  $\zetap'$ and its
inverse ${\zetap^*}\!\!'$ are depicted in Figure~\ref{fig-visco-plastic-2}. Note that we
can identify $\bm\sigma=\zetap'(\DTee)$ from Figure~\ref{fig-visco-plastic-2}-middle
by analyzing the total stress $\bm\sigma$ in the cases when the plastic
element is either active on the stress $\sigma_\text{\!\sc a}^{}$, i.e.\ 
$\bm\sigma=\sigma_\text{\!\sc a}^{}\DTee/\Mid\DTee\Mid+D_3^{}\DTee$, or inactive so that
$\bm\sigma=D_2^{}\DTee+D_3^{}\DTee$. The switching between these two regimes occurs
when $D_2^{}\DTee=\sigma_\text{\!\sc a}^{}\DTee/\Mid\DTee\Mid$, from which we obtain the
critical strain rate $\Mid\DTee\Mid=\sigma_\text{\!\sc a}^{}/D_2^{}$.

An alternative model from Figure~\ref{fig-visco-plastic-1}-right 
leads to the overall dissipation potential
\begin{align}\label{GSM-Jeffreys-plast+++}
\zetap=\big(\widetilde\zeta_1^{}{+}\widetilde\zeta_3^{}\big)\,\Box\,\widetilde\zeta_2^{}
\ \ \text{ with }\ 
\widetilde\zeta_1^{}(\cdot)=\widetilde\sigma_\text{\!\sc a}^{}\Mid\cdot\Mid
\ \text{ and }\ \widetilde\zeta_i^{}(\cdot)=\Frac12\widetilde D_i^{}\Mid\cdot\Mid^2,\ \ i=2,3.
\end{align}
Now, $\zetap$ is the Yosida approximation of the curve from Fig.\,\ref{fig-plastic},
specifically $\mathscr{Y}_{1/\widetilde D_2}(\widetilde\zeta_1^{}{+}\widetilde\zeta_3^{})$
with $\mathscr{Y}$ from \eq{Yosida-approx} below.
In fact, \eq{GSM-Jeffreys-plast+++} is equivalent with \eq{GSM-Jeffreys-plast++} when taken
\begin{align}
\widetilde\sigma_\text{\!\sc a}^{}=\sigma_\text{\!\sc a}^{}\Big(1{+}\Frac{D_3}{D_2}\Big)\,,
\ \ \ \ 
\widetilde D_2^{}=D_2^{}+D_3^{}\,,\ \text{ and }\ \widetilde D_3^{}=D_3^{}\Big(1{+}\Frac{D_3}{D_2}\Big)\,.
\label{visco-plastic-comparison}\end{align}
More in detail, analyzing the mode from Fig.\,\ref{fig-visco-plastic-1}-right
in the situation that the plastic element is sliding gives
$\bm\sigma=\widetilde\sigma_\text{\!\sc a}^{}\DTee_1^{}/\Mid\DTee_1^{}\Mid
+\widetilde D_3^{}\DTee_1^{}=\widetilde D_2^{}\DTee_2^{}$ with
$\DTee=\DTee_1^{}{+}\DTee_2^{}$, so that
\begin{align}
\bm\sigma=\widetilde\sigma_\text{\!\sc a}^{}
\frac{\widetilde D_2^{}}{\widetilde D_2^{}{+}\widetilde D_3^{}}
\,\frac{\DTee}{\Mid\DTee\Mid}+
\frac{\widetilde D_2^{}\widetilde D_3^{}}{\widetilde D_2^{}{+}\widetilde D_3^{}}\DTee\,.
\end{align}
When  the plastic element is not activated, $\bm\sigma=\widetilde D_2^{}\DTee$.
The switching between these two regimes occurs if
$\Mid\widetilde D_2^{}\DTee\Mid=\widetilde\sigma_\text{\!\sc a}^{}$. This gives
\eq{visco-plastic-comparison} by comparing it with Figure~\ref{fig-visco-plastic-2}-middle.

Thus, both variants in Figure~\ref{fig-visco-plastic-1} are mutually equivalent
and the effective viscosity now looks as
\begin{align}\nonumber\\[-2.7em]\label{effect-visco-1}
\mu_{\rm eff}^{}(\DTee)=\min\Big(\Frac{\sigma_\text{\!\sc a}^{}}{\Mid\DTee\Mid}\,,\,D_2^{}\Big)+D_3^{}\,.
\end{align}

When $\zeta_3^{}=\widetilde\zeta_3^{}$ vanishes (i.e.\ $D_3=\widetilde D_3\to0$),
both \eq{GSM-Jeffreys-plast++} and \eq{GSM-Jeffreys-plast+++} approach the 
visco-perfectly-plastic 
model $\zetap=\zeta_1^{}\Box\,\zeta_2^{}$ or $\widetilde\zeta_1^{}\Box\,\widetilde\zeta_2^{}$
as used in \eq{GSM-Jeffreys-plast+} or \eq{GSM-Jeffreys-plast+++}, respectively.
To approach $\zetap=\zeta_1^{}\!+\zeta_3^{}$ or $\widetilde\zeta_1^{}\!+\widetilde\zeta_3^{}$
like used in \eq{GSM-Jeffreys-plast}, one should make $D_2\to\infty$ in 
\eq{GSM-Jeffreys-plast++} or $\widetilde D_2\to\infty$ in \eq{GSM-Jeffreys-plast+++}.

One can also consider some serial and parallel combination of four or more perfect-plastic
and linear-creep dissipative elements. For example, parallel combination of
$m$ visco-perfectly-plastic fluid elements from Figure~\ref{fig-visco-plastic}-right and one
 Newton  element as in Figure~\ref{fig-visco-plastic-1}-left leads to overall
dissipation potential $\zetap$ as
\begin{align}\nonumber
\zetap=\!\SUM_{i=1}^m(\zeta_{1,i}^{}\,\Box\,\zeta_{2,i}^{})+\zeta_3^{}
\ \ &\text{ with }\ 
\zeta_{1,i}^{}(\cdot)=\sigma_{\text{\!\sc a},i}^{}\Mid\cdot\Mid
\\[-.4em]\ &\text{ and }\ \ 
\zeta_{2,i}^{}(\cdot)=\tfrac12D_{2,i}^{}\Mid\cdot\Mid^2
\label{GSM-Jeffreys-plast-many}\end{align}
with $i=1,..., m$ and with $\zeta_3^{}$ as in \eq{GSM-Jeffreys-plast++}.
The corresponding effective viscosity gives a 
generalization of \eq{effect-visco-1} as
\begin{align}\nonumber\\[-2.5em]\label{effect-visco-2}
\mu_{\rm eff}^{}(\DTee)=\SUM_{i=1}^m
\min\Big(\Frac{\sigma_{\text{\!\sc a},i}^{}}{\Mid\DTee\Mid}\,,\,D_{2,i}^{}\Big)+D_3^{}\,.
\end{align}

Notably, the effective viscosity $\mu_{\rm eff}^{}=\mu_{\rm eff}^{}(\DTee)$
obtained by such type of models in \eq{effect-visco} or \eq{effect-visco-2}
is decreasing with $\Mid\DTee\Mid$ increasing. Such phenomenon is known
as {\it shear-thinning non-Newtonian fluids} and 
allows for a certain ``lubrication'' on bands between two bulk areas
moving with different velocities, being applied naturally only to
the deviatoric component of $\DTee$. Actually, such shear-thinning
lubrication-like phenomenon occurs in many application in engineering (e.g.\
modelling of polymers or blood flow) or in geophysics to model cold heavy
slabs descenting into warmer mantle and can be modelled 
by a general convex potential $\zetap$ with a sub-quadratic growth.
The counter-case is shear-thickening models, which needed convex potentials
$\zetap$ with a super-quadratic growth that does not have any
analog composed from the linear viscous and the perfectly plastic elements
like on Figures~\ref{fig-visco-plastic} or \ref{fig-visco-plastic-1}.

\begin{remark}[{\sl Serial linear viscosities.}]\label{rem-serial-visco}\upshape
The serial arrangement of linear viscosities governed by the potentials
$\zeta_i(\DTee_i)=\frac12D_i\Mid\DTee_i\Mid^2$
with the moduli $D_i$ for $i=1,...,m$ made through the split of the strain rate
\begin{align}\nonumber\\[-2.8em]
    \ \DTee=\SUM_{i=1}^m\DTee_i
\label{serial-visco-split}\end{align}
and results to the potential $\zetap=\zeta_1\Box\cdots\Box\,\zeta_m$. This
potential is again quadratic and gives the linear viscosity
$\zetap(\DTee)=\frac12D\Mid\DTee\Mid^2$ with the modulus $D$ as
the {\it harmonic mean}  multiplied by the factor $m$,  i.e.
\begin{align}\nonumber\\[-2.8em]
\mu_{\rm eff}^{}=D=\Big(\SUM_{i=1}^m\Frac1{D_i}\Big)^{-1}\,.
\label{serial-visco-harmonic}\end{align}
More in detail, realizing that $\zeta_i^*(\bm\sigma)=\frac12D_i^{-1}\Mid\bm\sigma\Mid^2$,
we have $\DTee={\zetap^*}{\!\!'}(\bm\sigma)=\sum_{i=1}^m\DTee_i$ with
$\DTee_i=D_i^{-1}\bm\sigma$ so that
$\zetap'(\DTee)=(\sum_{i=1}^mD_i^{-1})^{-1}\DTee=D\DTee$ with $D$
from \eq{serial-visco-harmonic}. 
\end{remark}

\begin{remark}[{\sl The harmonic mean general.}]\upshape
  Coming back to \eq{GSM-Jeffreys-plast+} for generally $m$ potentials
  $\zeta_i$ and using \eq{serial-visco-split}, we realize that
$\DTee_i={\zeta_i^*}'(\bm\sigma)$ for $i=1,...,m$. This is 
$\bm\sigma=\zeta_i'(\DTee_i)=\bm\mu_i(\DTee_i)\DTee_i$ or, in other words, 
  $\DTee_i=\bm\mu_i(\DTee_i)^{-1}\bm\sigma$ for the viscosity
  $\bm\mu_i(\DTee_i)=\zeta_i'(\DTee_i)\DTee_i^{-1}$ which, in general, depends on
  $\DTee_{ i}$  unless $\zeta_i$ would be quadratic like in
  Remark~\ref{rem-serial-visco}.  Then 
  \begin{align}\nonumber
   \ \DTee\!\!\stackrel{\eq{serial-visco-split}}{=}\SUM_{i=1}^m\DTee_i
    =\!\!\!\lineunder{\Big(\SUM_{i=1}^m\bm\mu_i(\DTee_i)^{-1}\Big)}
   {\qquad\qquad$=:\bm\mu_{\rm eff}^{}(\DTee)^{-1}$}\!\!\bm\sigma
    \end{align}
  and thus this  straightforward   procedure gives the effective viscosity
   as the harmonic mean multiplied by the factor $m$, i.e.
\begin{align}
  \bm\mu_{\rm eff}^{}(\DTee)=\Big(\SUM_{i=1}^m
\bm\mu_i(\DTee_i)^{-1}\Big)^{-1}\text{ with $\DTee$ from \eq{serial-visco-split}}\,.
\label{serial-visco-harmonic+}\end{align}
 Here, since  $\bm\mu$'s are matrix-valued, $(\cdot)^{-1}$
means the inverse matrix.  It is to be noted that \eq{serial-visco-harmonic+} is rather
implicit because the partition of $\DTee$ into the particular strain rates $\DTee_i$
from \eq{serial-visco-split} must be determined separately, like it was done in
Remark~\ref{rem-serial-visco} for the linear situation where simply 
$\mu_i=D_i$  were scalar-valued  and constant,  and then 
\eq{serial-visco-harmonic+} gives just \eq{serial-visco-harmonic}.
The combination via the harmonic mean \eq{serial-visco-harmonic+}
is commonly used,  albeit rather empirically  in a simplified way by
replacing all instances with $\DTee_i$
in \eq{serial-visco-harmonic+} by the total strain rate $\DTee$
to eliminate the mentioned difficulties with evaluation of
the partial strain rates $\DTee_i$, i.e.
\begin{align}
\bm\mu_{\rm eff}^{}(\DTee)=\Big(\SUM_{i=1}^m
\bm\mu_i(\DTee)^{-1}\Big)^{-1}\,.
\label{serial-visco-harmonic+simple}\end{align}
For usage of \eq{serial-visco-harmonic+simple}  in geophysical modelling
of the Earth's mantle dynamics, it phenomenologically involves specific
nonlinear viscosities, cf.\ e.g.\ in
\cite{BHBF22EPIR,CizBin13EMSI,CBSM12VELM,Gery19INGM,KaYuKa99TMEL,MCSC12EVTD,PoCiBe21FBSD,BeKeYu93ECNN}.
 This  will be discussed in the next Section~\ref{sec-nonlin-visco}.
 Noteworthy, 
in nonlinear cases when some (or each) $\bm\mu_i$ depends on
$\DTee_{ i}$, \eq{serial-visco-harmonic+simple}
 yields in general a different effective viscosity than
\eq{serial-visco-harmonic+}; recall that \eq{serial-visco-harmonic+}
 which is just \eq{GSM-Jeffreys-plast+} adapted
for the general $m$ potentials $\zeta_i$ 
gives $\bm\sigma=\zetap'(\DTee)$ with the effective viscosity
$\bm\mu_{\rm eff}^{}(\DTee)=\zetap'(\DTee)\DTee^{-1}$
with $\zetap=\zeta_1\Box\cdots\Box\,\zeta_m$. This is, in general,
 indeed  different from \eq{serial-visco-harmonic+simple}, as
will be shown in Remark~\ref{rem-dif-dis}  below.
\end{remark}

\begin{remark}[{\sl Alternative phenomenological combination of visco-plasticity.}]\label{rem-serial-visco+}\upshape
Applying the empirical serial-combination approach \eq{serial-visco-harmonic+} to the
parallel-serial rheology in Figure~\ref{fig-visco-plastic-1}-left would modify \eq{effect-visco-1} as
\begin{align}\label{effect-visco-1-mod}
\mu_{\rm eff}^{}(\DTee)=
\bigg(\Frac{\Mid\DTee\Mid}{\sigma_\text{\!\sc a}^{}}+\Frac{1}{D_2^{}}\bigg)^{-1}\!\!+D_3^{}
=\Frac{\sigma_\text{\!\sc a}^{}(D_2^{}{+}D_3^{})+D_2^{}D_3^{}\Mid\DTee\Mid}
{\sigma_\text{\!\sc a}^{}+D_2^{}\Mid\DTee\Mid}\,.
\end{align}
On the other hand, applying \eq{serial-visco-harmonic+}
to the serial-parallel rheology in Figure~\ref{fig-visco-plastic-1}-right would lead to
\begin{align}\nonumber\\[-2.5em]\label{effect-visco-2-mod}
\mu_{\rm eff}^{}(\DTee)=
\bigg(\Frac{1}{\widetilde\sigma_\text{\!\sc a}^{}/\Mid\DTee\Mid+\widetilde D_3^{}}
+\Frac{1}{\widetilde D_2^{}}\bigg)^{-1}\!\!
=\Frac{\widetilde\sigma_\text{\!\sc a}^{}\widetilde D_2^{}+
  \widetilde D_2^{}\widetilde D_3^{}\Mid\DTee\Mid}
{\widetilde\sigma_\text{\!\sc a}^{}+(\widetilde D_2^{}{+}\widetilde D_3^{})\Mid\DTee\Mid}\,,
\end{align}
as used e.g.\ in \cite{Tosi15CBVT}. Notably, these two variants are {\it not} equivalent
to each other, in contrast to the rigorous combination based on the convex analysis
leading to \eq{GSM-Jeffreys-plast++} and \eq{GSM-Jeffreys-plast+++} which
{\it are} mutually equivalent provided \eq{visco-plastic-comparison} holds. More in detail,
realize that, to obtain the same limits for $\Mid\DTee\Mid\to0$ and for
$\Mid\DTee\Mid\to\infty$, one would need $\widetilde D_2^{}=D_2^{}+D_3^{}$ and 
$D_3^{}=\widetilde D_2^{}\widetilde D_3^{}/(\widetilde D_2^{}{+}\widetilde D_3^{})$,
respectively. Yet,  it  cannot be fulfilled simultaneously except
$D_2^{}=\widetilde D_2^{}$, which proves that \eq{effect-visco-1-mod} and
\eq{effect-visco-2-mod} are indeed different. On the other hand, for
$D_3^{}=\widetilde D_3^{}=0$, both variants coincide with each other,
and are often used in literature, cf.\ e.g.\ \cite{RGNT15FRSL} and
\eq{example-Norton-Hoff} below for $n=\infty$.
\end{remark}

\section{Some nonlinear viscosities}\label{sec-nonlin-visco}

The convex potentials $\zetap$ as derived above, in particular
\eq{GSM-Jeffreys-plast++} and \eq{GSM-Jeffreys-plast-many}, should be
understood only as special cases of general phenomenological convex potentials.
Actually, the used two options of degree-2 and degree-1  positively 
homogeneous potentials of the linear-viscous or the perfectly-plastic ``bricks''
are rather borderline cases for low strain rates, modelling either
 non-activated (i.e.\ creep-type)  or activated 
(i.e.\ plastic-type)  dissipative process{es}, respectively.
In contrast to the above models composed from such elements, general convex
potentials can have possibly a sub-quadratic or a super-quadratic growth that
could model {\it shear-thinning} or {\it shear-thickening fluids}, respectively,
the latter one being sometimes used in engineering and also in geophysical
modeling of solid Earth for magma flow, cf.\ \cite{PCUC16RFLM}. The natural
simplest generalization of the linear viscosity is the {\it Norton-Hoff}
(sometimes called Norton-Bailey) {\it model} for (if used in the deviatoric part
alone without elasticity) so-called {\it power-law fluids}, 
governed by the phenomenological law
\begin{align}
\bm\sigma=D\Mid\DTee\Mid^{1/n-1}\DTee
\label{Norton-Hoff}\end{align}
with $n>0$ and with the viscosity-like coefficient $D$ in Pa${\cdot}$s$^{1/n}$.
The corresponding effective viscosity is then
$\mu_{\rm eff}^{}(\DTee)=D\Mid\DTee\Mid^{1/n-1}$ while
the dissipation potential is $\zetap(\DTee)=\frac{n}{n+1}D\Mid\DTee\Mid^{1+1/n}$.
Let us remark that \eq{Norton-Hoff} is sometimes written in the form
$\DTee=D'\Mid\bm\sigma\Mid^{m-1}\bm\sigma$, which is equivalent if $m=1/n$
and $D'=D^{1/n}$ while, in mathematical literature, the exponent
$1/n-1$ in \eq{Norton-Hoff} is rather written as $p-2$, which is obviously
equivalent to \eq{Norton-Hoff} if $p=1+1/n$. Notably, the convex conjugate
to the potential $\frac1{1+1/n}D\Mid\cdot\Mid^{1+1/n}$ governing \eq{Norton-Hoff}
is then $\frac1{(1+n)D^n}\Mid\cdot\Mid^{1+n}$. For $n=1$, this leads to the linear
viscosity as used in
Sections~\ref{sec-viscoplasticity}--\ref{sec-viscoplasticity-2},
sometimes under the name {\it diffusion creep} \cite{KaYuKa99TMEL}. 
If not combined with any plasticity, this is also called Newtonian fluids.
The exponent $n<1$ or $n>1$ leads to a particular class of the shear-thickening
or the shear-thinning (non-Newtonian) fluids, respectively, being sometimes
referred as {\it Ostwald–de Waele fluids} when $m$ instead of $1/n$ is written.
For $n>1$, this sort of nonlinear viscosity is used for rock dynamic in
the Earth's mantle.

Most common choice of the exponent $n\sim3-3.5$ is used to model so-called 
{\it dislocation} (or {\it ductile}) {\it creep} as e.g.\ in
\cite{BeKeYu93ECNN,Bill08MDSS,CBSM12VELM,KarWu93RUMS,KaYuKa99TMEL,PaCiPo24DCAL},
occasionally even $n\sim2.3 - 4.7$ as used in \cite{Buro11RSL,PoCiBe21FBSD}.
The same power-law fluid model is used for the creep of polycrystalline ice,
where it is referred as the {\it Glen} \cite{Glen55CPI} or {\it Glen–Nye
  law} \cite{Nye53FLMG}, with the most commonly
used $n=3\pm0.2$, cf.\ e.g.\ \cite{CufPat10PG,GolKoh01SDIE}.
Ignoring particular microscopical mechanisms, both belong to a broader class
of the {\it ductile creep}. For $n\to\infty$ in
\eq{Norton-Hoff}, this model approaches the dry-friction type (i.e.\ the
perfect plasticity) model with $D\to\sigma_\text{\!\sc a}^{}$ as used in
Figure~\ref{fig-visco-plastic}; note that the aforementioned physical unit
Pa${\cdot}$s$^{1/n}$=J${\cdot}$s$^{1/n}$/m$^3$ of $D$ approaches the physical
unit of the activation yield stress
$\sigma_\text{\!\sc a}^{}$, i.e.\ Pa=J/m$^3$. In reality, materials cannot
withstand too much large shear stress, for which viscosities with fast growing
conjugate potentials are used, e.g.\ of the Norton-Hoff power-law type under the
name {\it stress limiter} with high $n$, specifically
$n=5$ in \cite{CHBV02IRWY,MCSC12EVTD} or $n=10$ in
\cite{CizBin13EMSI,CBSM12VELM,PaCiPo24DCAL,PoCiBe21FBSD},
or even a super-power (exponential) growth in a Peierls'-type creep or just
(perfect) plastic creep, i.e.\ for $n=\infty$, as used e.g.\ in
\cite{BHBF22EPIR,Gler18NVAB,LiGeCo19VSSM,LXGB13CCCN}.

A serial combination of the perfect plasticity and the power-law viscosity,
i.e.\ a polynomial but nonquadratic $\zeta_2$ in the model from
Figure~\ref{fig-visco-plastic}-right, leads to the so-called
{\it Herschel–Bulkley fluid}, which is used to model magma flow, cf.\
\cite{ChPiHa19MVLF,FarHub23RSVF,KoChDi22MSR},
or for the general mantle rheology \cite{LXGB13CCCN}.
Combining the power-law creep
\eq{Norton-Hoff} with the perfectly plasticity (as e.g.\ in \cite{LiGeCo19VSSM})
in a way phenomenologically motivated by the min-formula in
\eq{effect-visco} with also the simplification like in
\eq{serial-visco-harmonic+simple}, we obtain the effective viscosity
\begin{align}\nonumber\\[-2.7em]
\mu_{\rm eff}^{}(\DTee)=\
\min\bigg(\Frac{\sigma_\text{\!\sc a}^{}}{\Mid\DTee\Mid}\,,\,\Frac{D}{\Mid\DTee\Mid^{1-1/n}}\bigg)\,.
\label{example-Herschel–Bulkley}\end{align}

Of course, the nonlinearly-viscous element governed by \eq{Norton-Hoff}
can be build into a visco-elastic rheology, possibly in the deviatoric and
the volumetric parts differently, specifically
the volumetric part uses solid-type models (often just ideally rigid, i.e.\
incompressible). To overcome the discontinuous behavior of $\bm\mu_{\rm eff}^{}$
at zero strain rate in the Bingham model, a so-called Papanastasiou
viscoplastic fluids governed  by 
$\bm\sigma=\sigma_\text{\!\sc a}^{}(1+c\Mid\DTee\Mid^n)^{1/n}\DTee/\Mid\DTee\Mid$
was devised in \cite{Papa87FMY}. Similar modification is the so-called
Casson fluid governed by $\bm\sigma=\sigma_\text{\!\sc a}^{}(1+c\Mid\DTee\Mid)^{1/2}\DTee/\Mid\DTee\Mid$.

Some other empirical viscoplastic models combine the harmonic mean for some
viscous power-law mechanisms \eq{example-Norton-Hoff} with the min-formula like
\eq{example-Herschel–Bulkley}, cf.\
\cite{Gler18NVAB,FisGer16EEPL,LWGB17MCDH,TiTaLo23TVVN} or other in parallel
(i.e.\ additively) with other power-law mechanisms \eq{example-Norton-Hoff},
the latter devised for improving the Glen model for the creep of the
polycrystaline ice \cite{GolKoh01SDIE,SoKaCa14WTPI}.

\begin{example}[{\sl Serial combination of the diffusive and the dislocation viscosities.}]\label{exa-dif-dis}\upshape
An explicit treatment of the infimal convolution in \eq{GSM-Jeffreys-plast+}
 can be  very nontrivial  (or even impossible)  in general, although
in some special case it is possible. Let us illustrate it on 
 the  serial combination of the diffusive and the dislocation viscosities, governed
by the potentials $\zeta_{\rm dif}(\DTee)=\frac12D_{\rm dif}\Mid\DTee\Mid^2$ and
$\zeta_{\rm dsl}(\DTee)=\frac n{n+1}D_{\rm dsl}\Mid\DTee\Mid^{1+1/n}$, respectively.
From the formula \eq{GSM-Jeffreys-plast+} above, we know the abstract formula
for  the resulting potential, namely $\zetap^{}=\zeta_{\rm dif}\,\Box\,\zeta_{\rm dsl}$.
It does not seem trivial to identify it explicitly for a general 
$n$ for which $\zetap^{}$ is the Yosida approximation
$\zetap^{}(\cdot)=\mathscr{Y}_{1/D_{\rm dif}}(\zeta_{\rm dsl})$. As said above
 in Remark~\ref{rem-serial-visco}, for $n=1$ we have simply the harmonic
mean $\zetap^{}=\frac12D_{\rm tot}^{}\Mid\DTee\Mid^2$ with
$D_{\rm tot}^{}=1/(1/D_{\rm dif}+1/D_{\rm dsl})$. In general, we can rely on that
$\zetap^*=\zeta_{\rm dif}^*+\zeta_{\rm dsl}^*$ with
$\zeta_{\rm dif}^*(\bm\sigma)=\frac1{2D_{\rm dif}}\Mid\bm\sigma\Mid^2$ and
$\zeta_{\rm dsl}^*(\bm\sigma)=\frac1{(n+1)D_{\rm dsl}^n}\Mid\bm\sigma\Mid^{1+n}$. Thus
\begin{align}
  \DTee={\zetap^*}\!\!'(\bm\sigma)={\zeta_{\rm dif}^*}\!\!\!\!'\:(\bm\sigma)+{\zeta_{\rm dsl}^*}\!\!\!\!'\:(\bm\sigma)
  =D_{\rm dif}^{-1}\bm\sigma+D_{\rm dsl}^{-n}\Mid\bm\sigma\Mid^{n-1}\bm\sigma\,.
\end{align}
Obtaining  explicitly the inverse which would express $\bm\sigma$ as a function
of $\DTee$ is not possible except the integer values of $n\le4$. For this,
taking into account the isotropy, one should solve the algebraic equation
\begin{align}\label{algebraic-eq}
D_{\rm dsl}^{-n}\sigma^n+D_{\rm dif}^{-1}\sigma-\varepsilon=0
\end{align}
for the scalar variables $\sigma\ge0$ and $\varepsilon\ge0$.
In particular, for $n=2$, we are to solve the quadratic algebraic equation,
which gives
\begin{align}\label{diffusive-dislocation2}
  \sigma=S(\varepsilon):=\sqrt{\frac{D_{\rm dsl}^4}{4D_{\rm dif}^2}+
  \varepsilon D_{\rm dsl}^2}-\frac{D_{\rm dsl}^2}{2D_{\rm dif}}\,.
\end{align}
It gives the stress
\begin{align}
\bm\sigma=\mu_{\rm eff}^{}(\DTee)\DTee\ \ \text{ with
the effective viscosity }\ \mu_{\rm eff}^{}(\DTee)=\Frac{S(\Mid\DTee\Mid)}{\Mid\DTee\Mid}\,.
\label{diffusive-dislocation2+}\end{align}
For $n=3$ which is a conventional choice for the dislocation viscosity, we are
to solve  \eq{algebraic-eq}, i.e.\ the cubic algebraic equation in the so-called
depressed form. This is to be solved by Cardano's formula: recall that this formula
gives the real root of the depressed cubic equation $x^3+px=q$ as
$x=\sqrt[3]{u_1}+\sqrt[3]{u_2}$ with $u_{1,2}=q/2\pm\sqrt{{q^2}/4+{p^3}/{27}}$.
This gives $\sigma=S(\varepsilon)$ now with
\begin{align}\label{diffusive-dislocation3}
   S(\varepsilon):=
  \sqrt[3]{\frac{\varepsilon}{2}D_{\rm dsl}^3\!+\!\sqrt{\frac{\varepsilon^2}{4}D_{\rm dsl}^6\!
      +\frac{D_{\rm dsl}^9}{\!27C_3^3D_{\rm dif}^3}}}
  +\sqrt[3]{\frac{\varepsilon}{2}D_{\rm dsl}^3\!-\!\sqrt{\frac{\varepsilon^2}{4}D_{\rm dsl}^6\!
      +\frac{D_{\rm dsl}^9}{\!27C_3^3D_{\rm dif}^3}}}
\end{align}
which is then to be used for \eq{diffusive-dislocation2+}.
Notably, for \eq{diffusive-dislocation3}, we have used the mentioned Cardano formula
with $p=D_{\rm dsl}^3/D_{\rm dif}$ and $q=\varepsilon D_{\rm dsl}^3$;
it should be emphasized that the cubic root is defined and used in
\eq{diffusive-dislocation3} also for negative arguments.
\end{example}

\begin{remark}[{\sl A comparison of \eq{serial-visco-harmonic+} and
      \eq{serial-visco-harmonic+simple}}.]\label{rem-dif-dis}\upshape
  In Figure~\ref{fig-diffusive-dislocation}, we compare
  \eq{diffusive-dislocation2}
  and \eq{diffusive-dislocation3} with the visco-perfect-plastic model
  from Figure~\ref{fig-plastic-in-series},
  and also with the conventionally used   phenomenologically simplified 
  formula 
  for the serial-like  arrangement of the various visco-plastic
  mechanisms  \eq{serial-visco-harmonic+simple},  here as
\begin{align}
\mu_{\rm eff}^{}(\DTee)=\frac{1}{1/D_{\rm dif}+\Mid\DTee\Mid^{1-1/n}/D_{\rm dsl}}\,.
\label{example-Norton-Hoff}\end{align}
Let us note that the exponent $1-1/n$ in \eq{example-Norton-Hoff} arises
from the potential
$\zeta:\DTee\mapsto\frac1{1{+}1/n}D_{\rm dsl}\Mid\DTee\Mid^{1+1/n}$ which leads
to the stress $\zeta'(\DTee)=D_{\rm dsl}\Mid\DTee\Mid^{1/n-1}\DTee$, cf.\
\eq{Norton-Hoff}, so that the inverse viscosity is
$\Mid\cdot\Mid^{1-1/n}/D_{\rm dsl}$ as occurs in \eq{example-Norton-Hoff} and
actually also in \eq{example-Herschel–Bulkley} before. 
It is used here for $n=2$ and 3, and for $n=\infty$, too. For
Figure~\ref{fig-diffusive-dislocation}, we
consider $D_{\rm dif}=1\,$Pa${\cdot}$s and $D_{\rm dsl}=1\,$Pa${\cdot}$s$^{1/n}$.
Notably,
for $n=\infty$, $D_{\rm dsl}$ is in the position of $\sigma_\text{\!\sc a}^{}$,
leading to the effective viscosity used in Remark~\ref{rem-serial-visco+}.
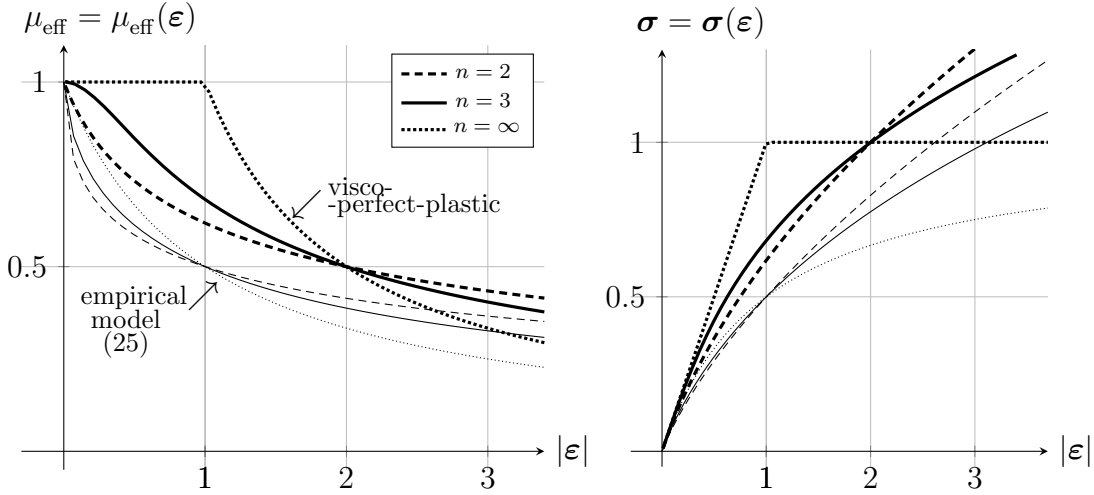
\begin{figure}[ht]\hspace*{.4em}
\begin{tikzpicture}  
    \begin{axis}[width=8.5 cm,height=7.2 cm,
        xmin=-0.3,xmax=3.4,ymin=-0.05,ymax=1.1,
        clip=true,  
        axis lines=center,  
        grid = major,  
        ytick={0, 0.5, 1},  
        xtick={0, 1, 1, 2, 3},  
        xlabel={$\Mid\DTee\Mid$}, ylabel={$\mu_{\rm eff}^{}=\mu_{\rm eff}^{}(\DTee)$\hspace*{-3em}},  
    every axis y label/.style={at=(current axis.above origin),anchor=south},  
    every axis x label/.style={at=(current axis.right of origin),anchor=west},  
      ]  
\addplot[very thick,domain=0.01:3.4,samples=70,densely dashed]{(sqrt(1/4+x)-1/2)/x};
\addplot[very thick,domain=0.01:3.4,samples=50]{(x/2+sqrt(x^2/4+1/27))^0.333/x-(sqrt(x^2/4+1/27)-x/2)^0.333/x};
\addplot[very thick,domain=0.01:3.4,samples=61,densely dotted]{min(1.0,1/x))};
\addplot[thin,domain=0:3.4,samples=50]{1/(1/1.0+x^0.66/1)};
\addplot[thin,domain=0:3.4,samples=50,densely dashed]{1/(1/1.0+x^0.5/1)};
\addplot[thin,domain=0:3.4,samples=50,densely dotted]{1/(1/1.0+x/1)};
\node at (axis cs:2.1,.717) {{\footnotesize visco-}};
\node at (axis cs:2.35,.67) {${\Large\swarrow}$\,{\footnotesize -perfect-plastic}};
\node at (axis cs:.61,.42) {{\footnotesize empirical}${\Large\nearrow}$ };
\node at (axis cs:.44,.36) {{\footnotesize model}};
\node at (axis cs:.44,.29) {{\footnotesize \eq{example-Norton-Hoff}}};
\legend{{\scriptsize $n=2\ $ },
       {\scriptsize $n=3\ $ },
        {\scriptsize $n=\infty$ },
              };
            \end{axis}  
  \end{tikzpicture}\hspace*{-.5em}
\begin{tikzpicture}  
    \begin{axis}[width=7.1 cm,height=7.1 cm,
        xmin=-0.3,xmax=3.7,ymin=-0.05,ymax=1.3,  
        clip=true,  
        axis lines=center,  
        grid = major,  
        ytick={0, 0.5, 1},  
        xtick={0, 1, 1, 2, 3},  
        xlabel={$\Mid\DTee\Mid$}, ylabel={$\bm\sigma=\bm\sigma(\DTee)$\hspace*{-2.5em}},  
    every axis y label/.style={at=(current axis.above origin),anchor=south},  
    every axis x label/.style={at=(current axis.right of origin),anchor=west},  
      ]  
\addplot[very thick,domain=0.01:3.4,samples=70,densely dashed]{(sqrt(1/4+x)-1/2)};
\addplot[very thick,domain=0.01:3.4,samples=50]{(x/2+sqrt(x^2/4+1/27))^0.333-(sqrt(x^2/4+1/27)-x/2)^0.333};
\addplot[very thick,domain=0.01:3.7,samples=61,densely dotted]{x*min(1.0,1/x))};
\addplot[thin,domain=0:3.7,samples=70]{x/(1/1.0+x^0.66/1)};
\addplot[thin,domain=0:3.7,samples=70,densely dashed]{x/(1/1.0+x^0.5/1)};
\addplot[thin,domain=0:3.7,samples=50,densely dotted]{x/(1/1.0+x/1)};
            \end{axis}  
\end{tikzpicture}\hspace*{-1em}
  \caption{
    {\sl A comparison of the serial arrangement of the linear viscous
      (i.e.\ diffusion) creep and the power-law Norton-Hoff model
      \eq{Norton-Hoff} with $n=2$, $n=3$ (i.e.\ the dislocation creep), and
      $n=\infty$ (i.e.\ the perfect plasticity as the min-formula in
      \eq{effect-visco}). Both the effective viscosity (left) and the
      corresponding stress (right) depending on the strain rate $\DTee$ are
      displayed. Also the harmonically-averaged empirical model
      \eq{example-Norton-Hoff} with the same coefficients for $n=2,3,$ and
      $\infty$ is depicted by the thin lines for a comparison.}
}
\label{fig-diffusive-dislocation}
\end{figure}
 Figure~\ref{fig-diffusive-dislocation} clearly shows
the different shape of functions resulting by
\eq{serial-visco-harmonic+} and \eq{serial-visco-harmonic+simple}
and thus it shows that these formulas lead to
different viscoplastic models even when some rescaling of
$\DTee$ and $\bm\sigma$ were applied.
\end{remark}

\begin{remark}[{\sl Generalized  Maxwell  rheologies.}]\label{rem-conjug-visco}\upshape
The formulas for effective visco\-sities in Example~\ref{exa-dif-dis} show that it 
 can be rather  nontrivial (or, for a general exponent $n$, even impossible)
to explicitly identify the dissipation potential $\zetap^{}$. However, if the
 serial  viscoplastic rheology  (which leads to the infimal
convolution due to \eq{GSM-Jeffreys-plast+})  is combined with elasticity
in series and the overall visco-elastodynamic system is formulated
suitably in terms of the strain rates, then only the conjugate potential
$\zetap^*$ occurs and the serial arrangement is reflected simply as a sum of
conjugate potentials, cf.\ \eq{GSM-Jeffreys-plast+}.
Here, it relies on the fact that the conjugate potentials can often be written
explicitly, thus avoiding the virtually difficult construction of the
infimal convolution. For the matrix-valued internal variable
$\ee_{\rm in}=\strain(\uu)-\ee_{\rm el}$ with
$\strain(\uu):=\frac12(\nabla\uu)^\top\!+\frac12\nabla\uu$ with $\uu$ the
(small) displacement, i.e.\ the {\it additive Green-Naghdi decomposition},
we then arrive to the visco-elastodynamic system for the couple
$(\vv,\ee_{\rm el})$: 
\begin{subequations}\label{GSM-small-strain-mod}
\begin{align}\label{GSM-small-strain1-mod}
  &\!\!\varrho\frac{\partial\vv}{\partial t}={\rm div}\,\bm\sigma
\ \ \ \text{ with }\ \ \bm\sigma=\varphi'(\ee_{\rm el})\ \ \text{ and }
\\[-.2em]&\label{GSM-small-strain3-mod}
\!\!\frac{\partial\ee_{\rm el}}{\partial t}=\DTee
-\SUM_{i=1}^m{\zeta_i^*}'(\bm\sigma)\ \ \text{ with }\
\DTee=\strain(\vv)\,;
\end{align}\end{subequations}
here we assumed the elasticity governed by a stored energy
$\varphi=\varphi(\ee_{\rm el})$. Of course, if some of the potentials $\zeta_i^{}$
is nonsmooth (as in the perfect plasticity), then ${\zeta_i^*}'$ is to be
understood as the subdifferential and the equality \eq{GSM-small-strain3-mod}
turns into an inclusion. The 
 Maxwell  rheology  (which standardly combines the elastic ``spring''
element with one Newtonian viscous element in series) 
is now generalized as depicted
in Figure~\ref{fig-rheology-Jeffreys-general}. For a serial combination with
the elastic solid \cite{BorDur20VRSS,DuBoPo19FTSB} or the viscoelastic
standard solid we refer e.g.\ to \cite{WQYZ14SREB}. Of course, a convected
variant of \eq{GSM-small-strain-mod} at large displacements exploiting objective
time derivatives works equally. This formulation is a particularly advantageous
formulation in the truly large-strain variant based on the {\it Kr\"oner-Lee-Liu
multiplicative} (in general not commutative)  {\it decomposition}
$\FF_{\rm el}^{}\FF_{\rm in}^{}$ of the deformation gradient instead of the
commutative Green-Naghdi additive decomposition in \eq{GSM-small-strain1-mod},
where $\bm\sigma$ is the {\it Cauchy stress} in \eq{GSM-small-strain1-mod}
given by $\varphi'(\FF_{\rm el}^{})\FF_{\rm el}^\top+\varphi'(\FF_{\rm el}^{})\bbI$
while
$\bm\sigma$ in \eq{GSM-small-strain3-mod} is then the (slightly different,
so-called) Mandel stress given as $\FF_{\rm el}^\top\varphi'(\FF_{\rm el}^{})$
acting equally in all visco-plastic elements $\zeta_i^*$ in the equation
$\frac{\partial}{\partial t}\FF_{\rm el}^{}=(\nabla\vv)\FF_{\rm el}^{}-(\vv{\cdot}\nabla)\FF_{\rm el}^{}
-\FF_{\rm el}^{}\Sum_{i=1}^m{\zeta_i^*}'(
\FF_{\rm el}^\top\varphi'(\FF_{\rm el}^{}))$  which is to replace
 \eq{GSM-small-strain3-mod}.
\begin{figure}
\begin{center}
\psfrag{e}{\small$\!\ee(\uu)$}
\psfrag{s}{\small$\bm\sigma$}
\psfrag{e1}{\small$\!\ee_{\rm el}^{}$}
\psfrag{e2}{\small$\ee_{\rm in}^{}$}
\psfrag{r}{\small$\varrho$}
\psfrag{c}{\small$\!\varphi'$}
\psfrag{d1}{\small$\bbD$}
\psfrag{zeta1}{\small$\zeta_1'$}
\psfrag{zeta2}{\small$\zeta_2'$}
\psfrag{zetam}{\small$\zeta_m'$}
\psfrag{s1}{\small$\bm\sigma_1^{}$}
\psfrag{s2}{\small$\bm\sigma_2^{}$}
\includegraphics[width=29em]{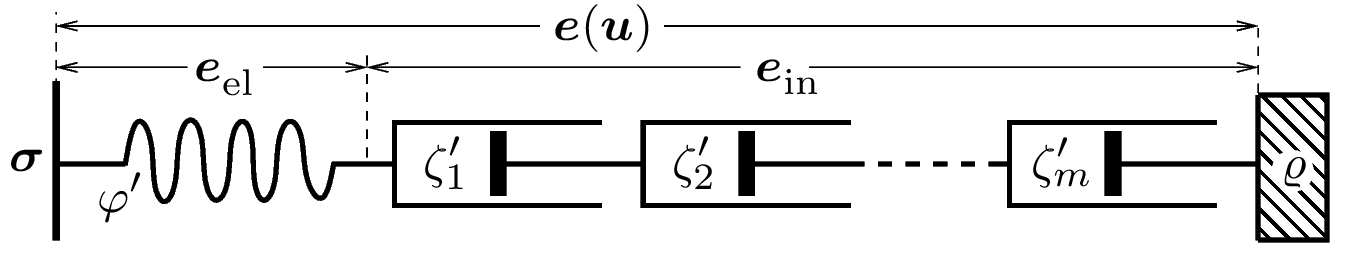}
\end{center}
\vspace*{-.8em}
\caption{A schematic 1-dimensional illustration of a generalized  Maxwell 
rheology with the stored energy $\varphi=
\varphi(\ee_{\rm el}^{})$ and the dissipation potential $\zetap=
\zetap(\DTee_{\rm in})$ for $\DTee_{\rm in}=\frac{\partial}{\partial t}\ee_{\rm in}$
with the viscoplastic potential $\zetap=\zeta_1^{}\Box\,\cdots\,\Box\,\zeta_m^{}$, cf.\
\eq{infimal-convolution-m} below.}
\label{fig-rheology-Jeffreys-general}
\end{figure}
\end{remark}

\section{Conclusion}\label{sec-conclude}

Classical models of viscoplasticity, expressed as various combinations
of (perfectly) plastic and (linear) viscous components, may arise to
a single dissipation potential which can be rigorously identified
using convex-analysis tools. The same applies also to various nonlinear
viscosities. Yet, it may be pretty nontrivial to identify
such viscoplastic potentials explicitly even in quite simple cases.
This is illustrated in particular power-law cases, as also routinely
used in geophysics for modelling of mantle dynamics and glaciology.

Having a single  dissipation potential at disposal may be advantageous
particularly for large-strain models. In these models, the related
inelastic distortions that describe the serial combination of the
involved viscoelastic or plastic models do not need to commute with
each other and attempting to write the model in terms of the
multiplicative decomposition of these distortions would create
an undesired ambiguity.

Of course, applicable viscoplastic models are incorporated into a broader
thermomechanical context, possibly including elasticity and some internal variables.
In particular, the coefficients in the above models typically depend
on temperature and volume, cf.\ e.g.\ \cite{BeKeYu93ECNN,CBSM12VELM,MCSC12EVTD,PoCiBe21FBSD,SoKaCa14WTPI},
as well as other variables, such as damage, varying medium composition,
a state variable in rate-and-state friction models, etc.

\section{Appendix: basics from convex analysis}\label{sec-append}

\def\In{{\,\in\,}}
  
Here we recall few standard definitions and formulas from convex analysis,
cf.\ e.g.\ the monographs \cite{EkeTem76CAVP,MorMaN23EPCA,Rock70CA}, confining
ourselves to the finite-dimensional space $\R^d$ with the Euclidean norm
$\Mid\cdot\Mid$ and with the duality (i.e.\ inner product) defined as
$a{\cdot}b:=\sum_{i=1}^da_ib_i$. Actually, the previous sections have used rather
the set of symmetric matrices in the position of $\R^d$.
For a convex lower semicontinuous $f:\R^d\to\R\cup\{\infty\}$, we define
the \emph{subdifferential} as a convex closed subset of the dual space to
$\R^d$, which is here again $\R^d$, namely
\begin{align}\label{app:subdifferential}
\partial f(v):=\big\{\sigma\In\R^d;\ \forall\, \widetilde v\In\R^d{:}\ \ 
f(v{+}\widetilde v) \geq f(v)+\sigma{\cdot} v\big\}.
\end{align}
When $f$ is differentiable with $f'$ denoting its derivative,
$\partial f$ is single-valued and $\partial f=\{f'\}$. The
\emph{convex conjugate} $f^*:\R^d\to\R\cup\{\infty\}$ of $f$ is defined via
\begin{align}\label{app:Legendre-Fenchel}
  f^*(\sigma):=\sup\big\{\sigma{\cdot} v -f(v);\ v\in\R^d\big\}.
\end{align}
Moreover, $f^*$ is called a \emph{convex conjugate} functional to $f$; note that
$f^*$, being the supremum of convex functions, is always convex. The so-called
Fenchel equivalences for subdifferentials read
\begin{equation}\label{e4.Fenchel-equiv}
\sigma \In \pl f(v) \quad \Leftrightarrow \quad v\In \pl f^*(\sigma)
\quad \Leftrightarrow \quad f(v)+f^*(\sigma)=\sigma{\cdot} v\,.
\end{equation}
This means
\begin{equation}
[\pl f]^{-1}=\pl f^*.
\end{equation}

Moreover, the {\it infimal convolution} of two
functions $f$ and $g$, denoted by $f\,\Box\,g$, is defined as
\begin{align}\label{infimal-convolution}
  \big[f\,\Box\,g\big](v):=\inf\big\{f(\tilde v)+g(v{-}\tilde v);\ \tilde v\in\R^d\big\}\,.
\end{align}
This ``operation'' between convex functions leads to a convex function and is
commutative and associative, in particular
\begin{align}\label{infimal-convolution-m}
\big[f_1\,\Box\,\cdots\,\Box\,f_m\big](v):=
\inf\Big\{\SUM_{i=1}^mf_i(v_i);\ \SUM_{i=1}^mv_i=v,\ v_i\In\R^d\Big\}\,.
\end{align}
This construction is useful for the following formula:
\begin{align}\label{infimal-convolution-conjugate}
\big[f\,\Box\,g\big]^*=f^*+g^*.
\end{align}

The infimal convolution with the function
$\frac{1}{2\varepsilon}\Mid\cdot\Mid^2$ gives
a generally applicable regularization construction called the
{\it Yosida} (sometimes refered as Moreau-Yosida) {\it ap\-pro\-ximation}, denoted as
$\mathscr{Y}_\varepsilon$, defined for a proper convex function
$f:\R^d\to\R\cup\{+\infty\}$ by 
\begin{align}\label{Yosida-approx}
\big[\mathscr{Y}_\varepsilon f\big](v):=\mbox{$\inf_{\tilde v\in\R^d}^{}$}f(\tilde v)
+\Frac1{2\varepsilon}\Mid\tilde v{-}v\Mid^2\,,
\end{align}
where $\varepsilon>0$ is a parameter.

\bigskip\bigskip

\noindent{\it Acknowledgments.}
Support from the CSF (Czech Science Foundation) project no.\,23-06220S and the
institutional support RVO: 61388998 (\v CR) is gratefully acknowledged.

{\small

} 



\begin{thebibliography}{10}

\vspace*{-.5em}\bibitem{ArGoHu17SSTZ}
R.~Agrusta, S.~Goes, and J.~{van Hunen}.
\newblock Subducting-slab transition-zone interaction: Stagnation, penetration
  and mode switches.
\newblock {\em Earth \& Planet. Sci. Letters}, 464:10--23, 2017.

\vspace*{-.5em}\bibitem{BecTeb12SFOM}
A.~Beck and M.~Teboulle.
\newblock Smoothing and ﬁrst order methods: a unified framework.
\newblock {\em SIAM J. Optim.}, 22:557--580, 2012.

\vspace*{-.5em}\bibitem{BHBF22EPIR}
W.M. Behr, A.F. Holt, T.W. Becker, and C.~Faccenna.
\newblock The effects of plate interface rheology on subduction kinematics and
  dynamics.
\newblock {\em Geophys. J. Int.}, 230:796--812, 2022.

\vspace*{-.5em}\bibitem{Bill08MDSS}
M.I. Billen.
\newblock Modeling the dynamics of subducting slabs.
\newblock {\em Annu. Rev. Earth Planet. Sci.}, 36:325--356, 2008.

\vspace*{-.5em}\bibitem{Buro11RSL}
E.B. Burov.
\newblock Rheology and strength of the lithosphere.
\newblock {\em Marine and Petroleum Geology}, 28:1402--1443, 2011.

\vspace*{-.5em}\bibitem{ChLaTh25ACHM}
F.~Cheng, R.~Lasarzik, and M.~Thomas.
\newblock {}{A}nalysis of a {C}ahn-{H}illiard model for viscoelastoplastic
  two-phase flows.
\newblock {\em {}WIAS preprint no.3247, Berlin}, 2025.
\newblock {}DOI: 10.20347/WIAS.PREPRINT.3247.

\vspace*{-.5em}\bibitem{ChPiHa19MVLF}
M.O. Chevrel, H.~Pinkerton, and A.J.L. Harris.
\newblock Measuring the viscosity of lava in the ﬁeld: A review.
\newblock {\em Earth-Sci. Rev.}, 196:Art.no.102852, 2019.

\vspace*{-.5em}\bibitem{CizBin13EMSI}
H.~{\v{C}\'{\i}\v{z}kov\'{a}} and C.R. Bina.
\newblock Effects of mantle and subduction-interface rheologies on slab
  stagnation and trench rollback.
\newblock {\em Earth \& Planetary Sci. Letters}, 379:95--103, 2013.

\vspace*{-.5em}\bibitem{CBSM12VELM}
H.~{\v{C}\'{\i}\v{z}kov\'{a}}, A.P. {van den Berg}, W.~Spakman, and C.~Matyska.
\newblock The viscosity of {E}arth's lower mantle inferred from sinking speed
  of subducted lithosphere.
\newblock {\em Phys. Earth $\&$ Planetary Interiors}, 200--201:56--62, 2012.

\vspace*{-.5em}\bibitem{CHBV02IRWY}
H.~{\v{C}\'{\i}\v{z}kov\'{a}}, J.~{van Hunen}, A.P. {van den Berg}, and N.J.
  Vlaar.
\newblock The inluence of rheological weakening and yield stress on the
  interaction of slabs with the 670 km discontinuity.
\newblock {\em Earth Planetary Sci. Letters}, 199:447--457, 2002.

\vspace*{-.5em}\bibitem{CufPat10PG}
K.M. Cuffey and W.S.B. Paterson.
\newblock {\em The Physics of Glaciers, {\rm 4th ed.}}
\newblock Elsevier, Amsterdam, 2010.

\vspace*{-.5em}\bibitem{BorDur20VRSS}
R.~{de Borst} and T.~Duretz.
\newblock On viscoplastic regularisation of strain-soften\-ing rocks and soils.
\newblock {\em Intl. J. Numer. Anal. Methods Geomech.}, 44:890--903, 2020.

\vspace*{-.5em}\bibitem{DuBoPo19FTSB}
T.~Duretz, R.~de~Borst, and L.~Le Pourhiet.
\newblock Finite thickness of shear bands in frictional viscoplasticity and
  implications for lithosphere dynamics.
\newblock {\em Geochemistry, Geophysics, Geosystems}, 20:5598--5616, 2019.

\vspace*{-.5em}\bibitem{EiHoMi22LHSV}
T.~Eiter, K.~Hopf, and A.~Mielke.
\newblock Leray-{H}opf solutions to a viscoelastoplastic fluid model with
  nonsmooth stress-strain relation.
\newblock {\em Nonlinear Analysis: Real World Applications}, 65:Art.no.103491,
  2022.

\vspace*{-.5em}\bibitem{EkeTem76CAVP}
I.~Ekeland and R.~Temam.
\newblock {\em Convex Analysis and Variational Problems}.
\newblock North Holland, Amsterdam, 1976.

\vspace*{-.5em}\bibitem{FarHub23RSVF}
S.A. Faroughi and C.~Huber.
\newblock Rheological state variables: A framework for viscosity
  parametrization in crystal-rich magmas.
\newblock {\em J. Volcanology \& Geothermal Res.}, 440:Art.no.107856, 2023.

\vspace*{-.5em}\bibitem{FisGer16EEPL}
R.~Fischer and T.~Gerya.
\newblock Early {E}arth plume-lid tectonics: A high-reso\-lu\-tion {3D}
  numerical modelling approach.
\newblock {\em J. Geodynam.}, 100:198--214, 2016.

\vspace*{-.5em}\bibitem{Gery19INGM}
T.V. Gerya.
\newblock {\em Introduction to Numerial Geodynamic Modelling, {\rm 2nd. ed.},}.
\newblock Cambridge Univ. Press, New York, 2019.

\vspace*{-.5em}\bibitem{Glen55CPI}
J.W. Glen.
\newblock The creep of polycrystalline ice.
\newblock {\em Proc. R. Soc. Lond. A}, 228:519--538, 1955.

\vspace*{-.5em}\bibitem{Gler18NVAB}
A.~{Glerum et al.}
\newblock Nonlinear viscoplasticity in {ASPECT}: benchmarking and applications
  to subduction.
\newblock {\em Solid Earth}, 9:267--294, 2018.

\vspace*{-.5em}\bibitem{GolKoh01SDIE}
D.L. Goldsby and D.L. Kohlstedt.
\newblock Superplastic deformation of ice: Experimental observations.
\newblock {\em J. Geophys. Res.: Solid Earth}, 106:11017--11030, 2001.

\vspace*{-.5em}\bibitem{Hube81RS}
P.~J. Huber.
\newblock {\em Robust Statistics}.
\newblock J.Wiley, New York, 1981.

\vspace*{-.5em}\bibitem{Huil15FMV}
R.R. Huilgol.
\newblock {\em Fluid Mechanics of Viscoplasticity}.
\newblock Springer, Berlin, 2015.

\vspace*{-.5em}\bibitem{JirBaz02IAS}
M.~Jir{\'a}sek and Z.P. Ba{\v{z}}ant.
\newblock {\em Inelastic Analysis of Structures}.
\newblock J.Wiley, Chichester, 2002.

\vspace*{-.5em}\bibitem{KaYuKa99TMEL}
M.~Kameyama, D.A. Yuen, and S.~Karato.
\newblock Thermal-mechanical effects of low-temperature plasticity (the
  {P}eierls mechanism) on the deformation of a viscoelastic shear zone.
\newblock {\em Earth \& Planetary Sci. Letters}, 168:159--172, 1999.

\vspace*{-.5em}\bibitem{Kara08DEMI}
S.~Karato.
\newblock {\em Deformation of Earth Materials -- An Introduction to the
  Rheology of Solid Earth}.
\newblock Cambridge Univ. Press, New York, 2008.

\vspace*{-.5em}\bibitem{KarWu93RUMS}
S.~Karato and P.~Wu.
\newblock Rheology of the upper mantle: a synthesis.
\newblock {\em Science}, 260:771--778, 1993.

\vspace*{-.5em}\bibitem{KoChDi22MSR}
S.~Kolzenburg, M.O. Chevrel, and D.B. Dingwell.
\newblock Magma / suspension rheology.
\newblock {\em Reviews in Mineralogy \& Geochemistry}, 87:639--720, 2022.

\vspace*{-.5em}\bibitem{LXGB13CCCN}
Z.-H. Li, Z.~Xu, T.~Gerya, and J.-P. Burg.
\newblock Collision of continental corner from {3-D} numerical modeling.
\newblock {\em Earth \& Planetary Sci. Lett.}, 380:98--111, 2013.

\vspace*{-.5em}\bibitem{LiGeCo19VSSM}
Z.H. Li, T.~Gerya, and J.A.D. Connolly.
\newblock Variability of subducting slab morphologies in the mantle transition
  zone: Insight from petrological-thermomechanical modeling.
\newblock {\em Earth-Sci. Rev.}, 196:Art.no.102874, 2019.

\vspace*{-.5em}\bibitem{LWGB17MCDH}
J.~Liao, Q.~Wang, T.~Gerya, and M.D. Ballmer.
\newblock Modeling craton destruction by hydration-induced weakening of the
  upper mantle.
\newblock {\em J. Geophys. Res. Solid Earth}, 122:7449--7466, 2017.

\vspace*{-.5em}\bibitem{MCSC12EVTD}
P.~{Maierov\'a et al.}
\newblock The effect of variable thermal diffusivity on kinematic models of
  subduction.
\newblock {\em J. Geophys. Res.}, 117:B07202, 2012.

\vspace*{-.5em}\bibitem{MorMaN23EPCA}
B.~Mordukhovich and N.~Mau Nam.
\newblock {\em An Easy Path to Convex Analysis and Applications, {\rm 2nd ed}.}
\newblock Springer, Cham/Switzerland, 2023.

\vspace*{-.5em}\bibitem{Nye53FLMG}
J.F. Nye.
\newblock The flow law of ice from measurements in glacier tunnels, laboratory
  experiments and the {J}ungfraufirn borehole experiment.
\newblock {\em Proc. Royal Soc. A}, 219:477--489, 1953.

\vspace*{-.5em}\bibitem{Papa87FMY}
T.C. Papanastasiou.
\newblock Flows of materials with yield.
\newblock {\em J. Rheol.}, 31:385--404, 1987.

\vspace*{-.5em}\bibitem{PaCiPo24DCAL}
V.~Pato\v{c}ka, H.~\v{C}\'{\i}\v{z}kov\'a, and J.~Pokorn\'y.
\newblock Dynamic component of the asthenosphere: Lateral viscosity variations
  due to dislocation creep at the base of oceanic plates.
\newblock {\em Geophys. Res. Letters}, 51:e2024GL109116, 2024.

\vspace*{-.5em}\bibitem{PCUC16RFLM}
M.~Pistone, B.~Cordonnier, P.~Ulmer, and L.~Caricchi.
\newblock Rheological flow laws for multiphase magmas: an empirical approach.
\newblock {\em J. Volcanology Geothermal Res.}, 321:158--170, 2016.

\vspace*{-.5em}\bibitem{PoCiBe21FBSD}
J.~Pokorn\'y, H.~\v{C}\'{\i}\v{z}kov\'a, and A.~van~den Berg.
\newblock Feedbacks between subduction dynamics and slab deformation: combined
  effects of nonlinear rheology of a weak decoupling layer and phase
  transitions.
\newblock {\em Phys. Earth \& Planetary Interiors}, 313:Art.no.106679, 2021.

\vspace*{-.5em}\bibitem{Rock70CA}
R.T. Rockafellar.
\newblock {\em Convex Analysis}.
\newblock Princeton Univ. Press, 1970.

\vspace*{-.5em}\bibitem{RGNT15FRSL}
A.~Rozel, G.J. Golabek, R.~N\"af, and P.J. Tackley.
\newblock Formation of ridges in a stable lithosphere in mantle convection
  models with a viscoplastic rheology.
\newblock {\em Geophys. Res. Lett.}, 42:4770--4777, 2015.

\vspace*{-.5em}\bibitem{SoKaCa14WTPI}
O.~Sou{\v{c}}ek, K.~Kalousov\'a, and O.~\v{C}adek.
\newblock Water transport in planetary ice shells by two-phase flow - a
  parametric study.
\newblock {\em Geophys. \& Astrophys. Fluid Dynamics}, 108:639--666, 2014.

\vspace*{-.5em}\bibitem{TiTaLo23TVVN}
J.~Tian, P.J. Tackley, and D.L. Lourenco.
\newblock The tectonics and volcanism of {V}enus: New modes facilitated by
  realistic crustal rheology and intrusive magmatism.
\newblock {\em Icarus}, 399:Art.no.115539, 2023.

\vspace*{-.5em}\bibitem{Tosi15CBVT}
N.~{Tosi at al.}
\newblock A community benchmark for viscoplastic thermal convection in a {2-D}
  square box.
\newblock {\em Geochem. Geophys. Geosyst.}, 16:2175--2196, 2015.

\vspace*{-.5em}\bibitem{BeKeYu93ECNN}
A.P. {van den Berg}, P.E. van Keken, and D.A. Yuen.
\newblock The effects of a composite non-{N}ewtonian and {N}ewtonian rheology
  on mantle convection.
\newblock {\em Geophys. J. Int.}, 115:62--78, 1993.

\vspace*{-.5em}\bibitem{WQYZ14SREB}
S.~Wang, J.~Qi, Z.~Yin, J.~Zhang, and W.~Ma.
\newblock A simple rheological element based creep model for frozen soils.
\newblock {\em Cold Regions Science and Technology}, 106–107:47--54, 2014.

\end{thebibliography}

\bigskip\bigskip\bigskip

\noindent
Mathematical Institute, Faculty of Math. \& Phys., Charles University,\\
Sokolovsk\'a 83, CZ-186~75~Praha~8,  Czech Republic,\\[.3em]
and\\[.3em]
Institute of Thermomechanics, Czech Academy of Sciences,\\
Dolej\v skova 5, CZ-18200~Praha~8, Czech Republic\\
email: ${\texttt{tomas.roubicek@mff.cuni.cz}}$

\end{document}